\documentclass{moriond}

\def\be{\begin{equation}}
\def\ee{\end{equation}}
\def\bea{\begin{eqnarray}}
\def\eea{\end{eqnarray}}

\begin{document}
\vspace*{4cm}
\title{THE SIMULATION OF DARKSIDE-20K CALIBRATION}

\author{MARIE VAN UFFELEN \\ on behalf of the DarkSide-20k collaboration}

\address{CPPM, Aix-Marseille Université, CNRS/IN2P3, Marseille cedex 09, France}

\maketitle\abstracts{
DarkSide-20k will be the next liquid argon TPC built to perform direct search for dark matter under the form of WIMPs. Its calibration to both signal and backgrounds is key as very few events are expected in WIMPs search. In the following proceeding, aspects of the calibration of the TPC of DarkSide-20k are presented: the calibration system itself, the simulations of the calibration programs and the simulations of the impact of the calibration system on the rest of the detector (reduction of the light collection efficiency in the veto buffer, induced background by the system in the TPC and veto).}

\section{Introduction}\label{Intro}

\begin{figure}
\begin{minipage}{0.5\linewidth}
\centerline{\includegraphics[width=8cm,keepaspectratio]{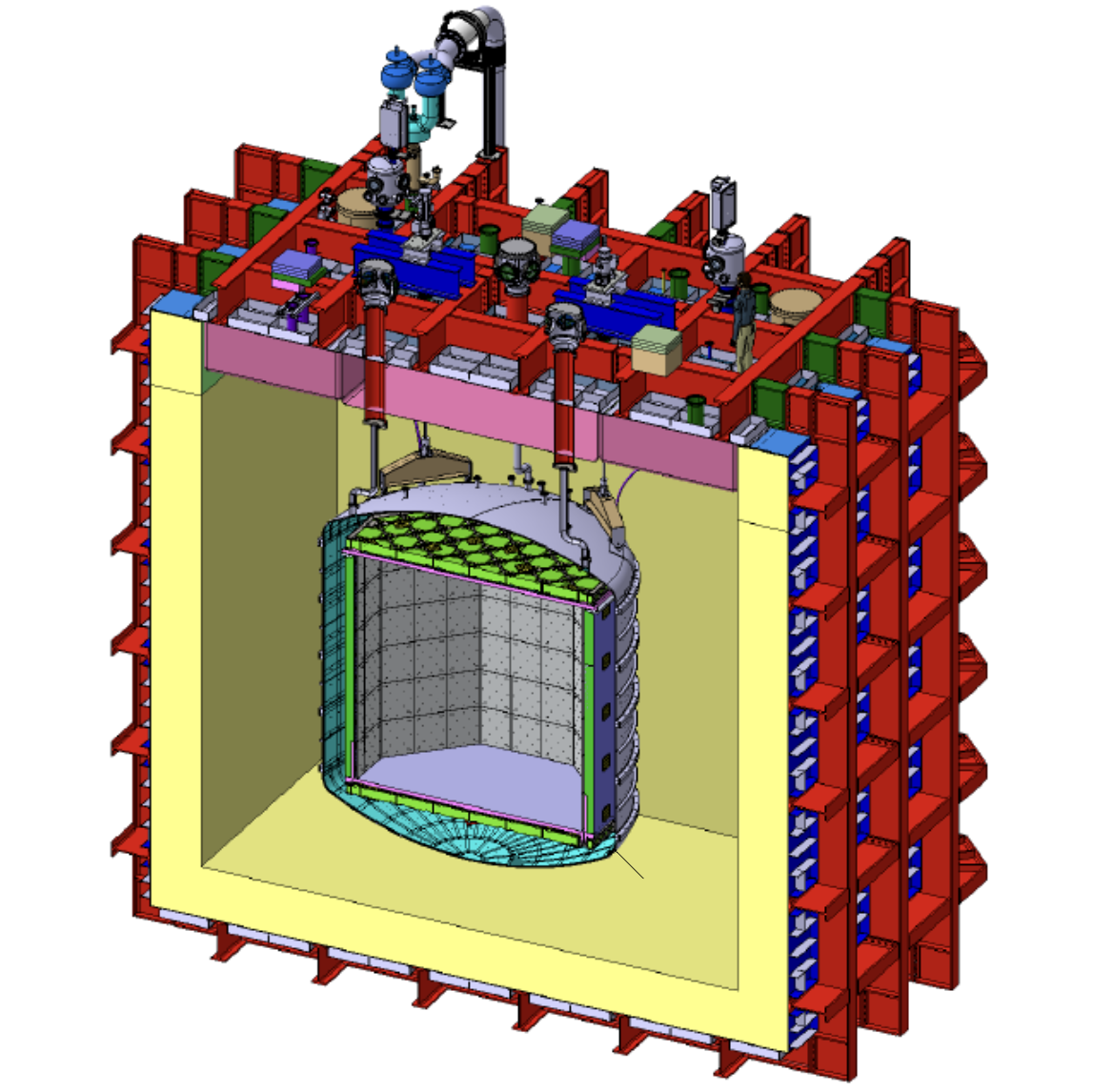}}
\end{minipage}
\hfill
\begin{minipage}{0.5\linewidth}
\centerline{\includegraphics[width=8cm,keepaspectratio]{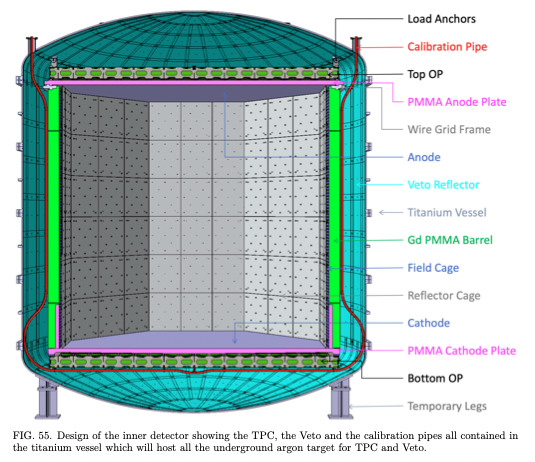}}
\end{minipage}
\hfill
\caption[]{Left: Cut view of the DarkSide-20k experiment, from the cryostat (outer part) to the TPC (inner part). The cryostat is a cube of 8.5~m side, the TPC is a 3.5~m high hexagon. Right: Cut view of the DarkSide-20k inner detector. The outer detector is the veto buffer, aiming at tagging background before it enters the TPC and interacts inside it. The inner part is the TPC, the detetor where the WIMP search will be performed. The calibration pipe (shown in red here) are located outside the TPC and inside the veto. This pipe exits the veto at its top cap and exits the detector by the top of the cryostat.}
\label{figCAO}
\end{figure}

After the successes of DEAP-3600~\cite{DEAP} and DarkSide-50~\cite{DS50} experiments, demonstrating the usability of Liquid Argon (LAr) Time Projection Chambers (TPC) for dark matter search, the next argon experiment built to detect WIMP dark matter will be DarkSide-20k~\cite{DS20k}, which will use a double phase TPC. It will be hold in the Hall C of the Gran Sasso National Laboratory (LNGS), in Italy, like its predecessor DarkSide-50. Its construction has started and it is expected to start operation end 2026. DarkSide-20k will be a layer detector working at cryogenic temperatures using LAr ($T_{boil}$~$=$~86~K). The inner layer will the inner detector (4.65~m high oval volume, filled with 100t of purified liquid underground argon), composed of two sub-detectors: a TPC at its center (3.5~m high hexagon) where the WIMP search will be performed, and a veto buffer around the latter (in order to tag residual background before it enters the active search volume). The inner detector is immersed in a large volume of liquid atmospheric argon, aiming at shielding against residual radioactivity and cosmic rays, and act as outer veto tagging. This is housed inside a cubic cryostat (8.5~x~8.5~x~8.5~$m^3$). Figure~\ref{figCAO}~-~left shows the CAD of DarkSide-20k detector, and Figure~\ref{figCAO}~-~right focuses on the inner detector.\\

Noble liquid double phase TPCs currently lead the direct search of WIMP dark matter~\cite{ESPP} thanks to their good signal - background rejection enabled by the two signals collected after an interaction in the detection volume. First, scintillation signal is emitted, the photons propagate quasi instantaneously in the detector and reach photo - detectors (in the case of DarkSide - 20k: SiPMs arrays at the top and bottom of the TPC), this signal is called S1. Second, there is ionization resulting from the interaction. Electrons can then drift through the TPC thanks to an electric field, until a thin layer of gaseous argon placed under a strong extraction electric field. There, the electrons produce photons via electroluminescence and are read by the same photo - detectors. This second - delayed - photon signal coming from ionisation electrons is called S2. The number of S1 and S2 photons released after an interaction is different in the case of a NR or an electronic recoil (ER), allowing TPCs to differentiate the two interactions, thus to distinguish the interaction of detector material with light particles (photons, electrons) or with heavy particles.\\
As one can expect from the low range of cross sections rejected by dark matter direct detection experiment (next generation experiments should be able to reject signal down to $\approx$~10$^{-48}$~cm$^2$ at 100~GeV~\cite{ESPP}), dark matter search expects very few WIMP events. In this scope, separation between signal and background is key. The main source of background in dark matter search come from radioactivity of the detector material, producing electrons, photons (these two are the most numerous) and neutrons. In the preparation of dark matter direct detectors, lots of efforts are put to minimize this radio contamination. Yet, there will be residual radioactivity that has to be understood, thus the use of double phase TPCs to discriminate main background (electrons and photons, ER producers) from signal (producing NR) thanks to the S1/S2 ratio. In addition, LAr has an additional asset precisely measured by DEAP-3600: the Pulse Shape Discrimination (PSD) which allows another $10^8$~to~$10^9$ rejection between ER and NR~\cite{DEAP}. The last remaining backgrounds are neutrons and neutrinos (controlled to be as low as possible, e.g. in the case of DarkSide-20k, the neutron's NR background budget is 0.1~events~/~10~years of data taking). \\
The calibration of DarkSide-20k is key in the scope of NR and ER characterization allowing background rejection. Plus, the energy and position resolution in the detector and its time stability have to be tested during the several calibration programs along the decade of data taking. Simulations were performed in order to optimize the calibration programs, from their duration to the impact of the calibration system on the detector. In the following proceeding, the calibration system of DarkSide-20k, the simulation of the calibration and the simulation of the impact of the system on the detector are presented. 

\section{The calibration of DarkSide-20k's TPC}\label{sec:Calib}

\begin{figure}
\begin{minipage}{0.5\linewidth}
\centerline{\includegraphics[width=7.5cm,keepaspectratio]{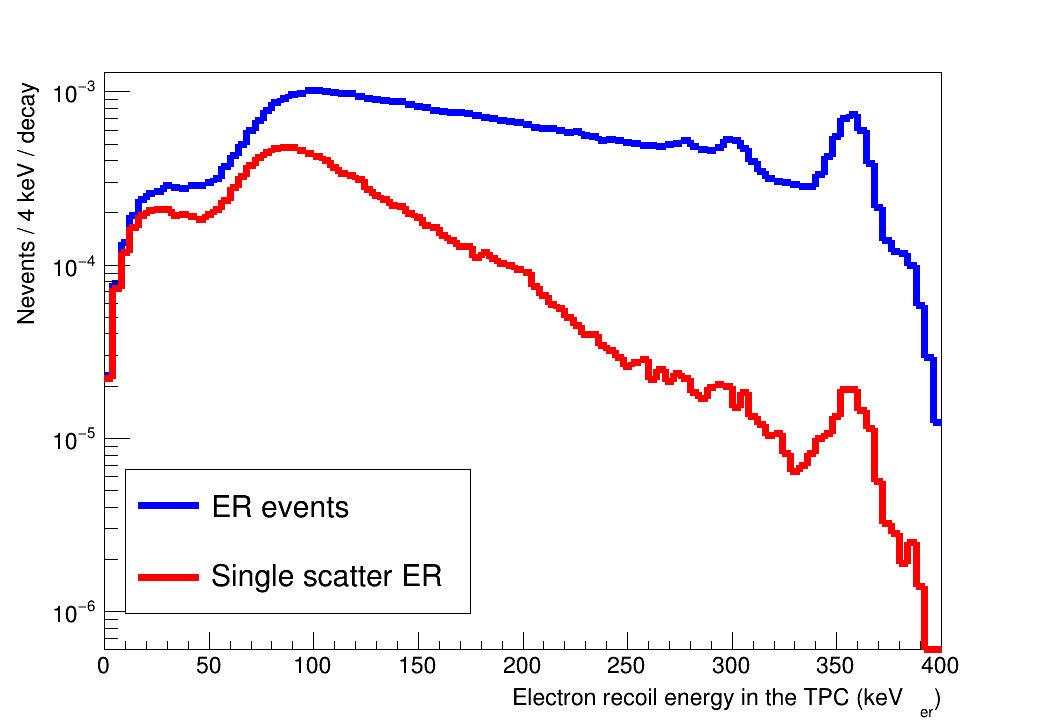}}
\end{minipage}
\hfill
\begin{minipage}{0.5\linewidth}
\centerline{\includegraphics[width=7.5cm,keepaspectratio]{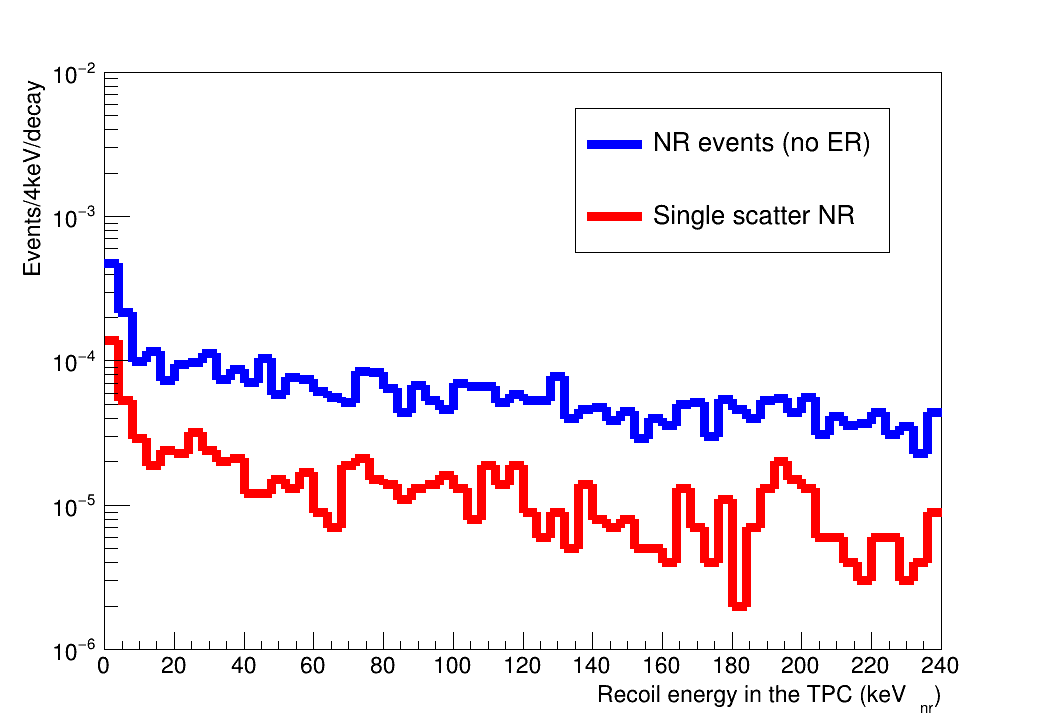}}
\end{minipage}
\hfill
\caption[]{Two examples of energy spectra obtained after the exposure of photon source (here $^{133}Ba$, left) and neutron source (here AmBe, right) on the side of the TPC.}
\label{figRates}
\end{figure}

The calibration of DarkSide-20k will be done two ways. First diffuse radioactive sources ($^{83m}$Kr, $^{220}$Rn) with short decay chain and lifetime will be inserted in the volume of the TPC together with residual radioactive $^{39}Ar$, this will allow uniform volume calibration. Yet, they can only provide ER calibration over a limited energy range. In order to perform the NR calibration, the collaboration is designing a system allowing to bring radioactive sources near the TPC, inside the cryostat and the inner detector. This system is a pair of tubes shaped to fit around the TPC (3~cm next to the TPC walls, 1~cm under the read-out system), dived inside the veto buffer and exiting the detector at the top of its cryostat. There, two pairs of motorized systems (one pair per tube) will circulate a rope in the tubes, to which the radioactive source will be attached. One calibration pipe can be seen on Figure~\ref{figCAO}~-~right. \\
The design of the TPC is thought to be able to reject at most the background, especially the remaining neutron background (the most dangerous as it mimicks the WIMP), using thick Gadollinium-loaded walls tagging neutrons with accompanying photons in the TPC. This property challenges the calibration that has to be carefully simulated to estimate its efficiency and duration, therefore its feasibility. \\
The simulation of the TPC calibration was performed using a GEANT4-based software. In this study, there are sources of neutrons: AmBe and AmC, and photons: $^{57}Co$, $^{133}Ba$, $^{22}Na$, $^{137}Cs$ and $^{60}Co$. The calibration using photons aims at understanding the ER background. Calibration with neutrons sources aims at producing NR, thus understanding both the background coming from neutrons and the WIMP signal. The energy spectra -such as the ones presented in Figure~\ref{figRates}- resulting from the interaction of emitted particles and the argon of the TPC allow to compute the rate of events in the TPC. In the left plot of Figure~\ref{figRates}, the blue histogram represents all ER events per decay in the TPC, and all NR events (no ER) in the right plot of Figure~\ref{figRates}. The red histograms are after the pure ER (for photon sources, left) or pure NR (for neutron sources, right) single scatter selection (meaning one interaction inside the TPC: the WIMP signature). The time needed to perform the calibration is computed thanks to the rate of such single scatters events, the activity of the source (tuned thanks to simulation to fit readout requirements) and the statistics needed for suitable calibration performance. The ER calibration and NR calibration should respectively last about one week (one day with certain calibration assumptions) and two weeks. \\

\section{Impact of the tubes on the inner detector performance}
As the tubes are dived inside the veto buffer and positioned very close to the TPC, they might degrade the performance of the inner detector, either by inducing background in both sub-detectors or by decreasing the light collection efficiency of the veto buffer. Thus, two sets of simulations were performed in order to study the influence of the tubes on the rest of the detector. \\
Even if the radioactivity of each detector's components material is controlled, there will be remaining radioactivity from the tubes. The tubes will be in stainless steel, which is contaminated by five isotopes: $^{238}U$, $^{235}U$, $^{232}Th$, $^{40}K$, $^{60}Co$ and $^{137}Cs$. All these radio-contaminants produce photons, the three first also produce neutrons. The simulations consisted in the uniform emission of photons and neutrons from the tubes with energy and activity as if they resulted from decays of the latter isotopes. In the analysis, pure ER (resp. pure NR) single scatters in the fiducial volume, in the energy region of interest and without veto tagging were selected for the ER (resp. NR) background rate estimation. The simulations lead to the conclusion that the tubes should induce less than 0.01~$\%$ of the NR background budget of DarkSide-20k. The ER background induced by the system should also be fully negligible, and will be ruled out by the PSD and the S1/S2 ratio anyway. 
\\

To finish on the simulation aspects, it is necessary to check the impact of the tubes on the veto's Light Collection Efficiency (LCE). Indeed, the veto will also collect photons from scintillation thanks to photo-detectors placed on its walls. The presence of the tubes can lower the veto's LCE or induce an asymmetry by absorbing or badly reflecting photons on their surface. In order to decrease as much as possible the effects of the tubes on the veto, different optical boundaries were tested, the best one is reflector (ESR foils) wrapped tubes. Furthermore, the LCE simulations showed a very small asymmetry in the veto induced by the pipes (less than 0.3~$\%$) and a relative loss of efficiency of 1$\%$ in average if the tubes are wrapped with a reflector foil.\\

\section{Conclusion}
Thanks to GEANT4-based simulations, the calibration procedure of DarkSide-20k could be established: the external calibration programs should last between one week (ER) and two weeks (NR), the activity of the source can be tuned to fit the detector requirements on pile-up. Plus, with two other sets of calibration, the low impact of the tubes on the inner detector (in terms of ER and NR backgrounds and diminution of veto LCE) could be affirmed.


\section*{Acknowledgements}
This work received support from the French government under the France 2030 investment plan, as part of the Excellence Initiative of Aix-Marseille University -- A*MIDEX (AMX-19-IET-008 -- IPhU).

\end{document}